\documentclass[11pt]{article}
\usepackage{cite}
\usepackage{geometry}                % See geometry.pdf to learn the layout options. There are lots.
\usepackage{graphicx}
\usepackage{amssymb}
\usepackage{amsmath,amssymb}
\usepackage{units}
\usepackage{epsfig}
\usepackage{graphicx}
\usepackage[usenames,dvipsnames]{color}

\def\byuk{{\bf Y}}

\title{
\vspace*{-2cm}
\begin{flushright}
\normalsize{ANL-HEP-PR-11-48\\
EFI-11-31\\
NUHEP-TH/11-21}
~\\
\end{flushright}
\vspace*{1.5cm}
An Alternative Yukawa Unified SUSY Scenario \\
\author{\textbf{James S. Gainer$^{a,b}$, Ran Huo$^c$ and Carlos E.M. Wagner$^{a,c,d}$} \\
~\\
\normalsize\emph{$^a$HEP Division, Argonne National Laboratory, 9700 Cass Ave., Argonne, IL 60439}\\
\normalsize\emph{$^b$Department of Physics and Astronomy, Northwestern University,  Evanston, IL 60208}\\
\normalsize\emph{$^c$Enrico Fermi Institute \& $^d$Kavli Institute for Cosmological Physics,}\\
\normalsize\emph{University of Chicago, Chicago, IL 60637} \\
}}
\begin{document}
\maketitle
\vspace*{0.5cm}
\begin{abstract}
Supersymmetric $SO(10)$ Grand Unified Theories with Yukawa unification represent an appealing possibility for physics beyond the Standard Model. However Yukawa unification is made difficult by large threshold corrections to the bottom mass. Generally one is led to consider models where the sfermion masses are large in order to suppress these corrections.  Here we present another possibility,
in which the top and bottom GUT scale Yukawa couplings are equal to a component of the charged lepton Yukawa matrix at the GUT scale in a basis where this matrix is not diagonal. Physically, this weak eigenstate Yukawa unification scenario corresponds to the case where the charged leptons that are in the $\boldsymbol{16}$ of $SO(10)$ containing the top and bottom quarks mix with their counterparts
in another $SO(10)$ multiplet. Diagonalizing the resulting Yukawa matrix introduces mixings in the neutrino sector. Specifically we find that for a large region of parameter space with relatively light sparticles, and which has not been ruled out by current LHC or other data, the mixing induced in the neutrino sector is such that $\sin^2 2 \theta_{23} \approx 1$, in agreement with data.
The phenomenological implications are analyzed in some detail.
\end{abstract}
\thispagestyle{empty}
\newpage

\section{Introduction}\label{introduction}

The idea of unification has been central to the development of modern physics. Supersymmetric (SUSY) Grand Unified Theories (GUTs) unify the gauge groups of the standard model as well as fermions and
bosons, in addition to potentially providing explanations of a variety of theoretical and observational problems or phenomena~\cite{Mohapatra:1999vv, Raby:2004px}. The choice of $SO(10)$ allows for the unification of all matter superfields (of each generation) into a 16-dimensional spinor representation, provided we add a superfield containing a right-handed neutrino to complete the multiplet~\cite{Georgi:1974my, Fritzsch:1974nn, GellMann:1976pg}. An advantage of including right-handed neutrinos is that it allows for the natural incorporation of the see-saw mechanism to
generate neutrino masses~\cite{Minkowski:1977sc, GellMann:1980vs,Yanagida:1979as, Yanagida:1980xy, Mohapatra:1979ia}. Finally, in the simplest $SO(10)$ SUSY GUT models one has Yukawa unification, by which one means that e.g.
\begin{equation}
  y_t = y_b = y_\tau,
\end{equation}
where $y_t$, $y_b$, and $y_\tau$ are the top, bottom, and tau Yukawa couplings at the GUT scale. (Simple Yukawa unification is not phenomenologically viable for the first and second generation.)

A great deal of work has gone into studying Yukawa unification (see for example \cite{Banks:1987iu,Olechowski:1988gh,Giudice:1988za,Ananthanarayan:1991xp, Anderson:1992ba,Ananthanarayan:1992cd,Anderson:1993fe,Barger:1993gh,Ananthanarayan:1994qt,Rattazzi:1995gk,Blazek:1996yv,Blazek:1996wa,Blazek:1997cs,Blazek:1999ue,Baer:1999mc,Gomez:1999dk,
Blazek:1999hz,Baer:2000jj,Baer:2001yy,Blazek:2001sb,Chattopadhyay:2001va,Blazek:2002ta,Gomez:2002tj,Tobe:2003bc,Gomez:2003cu,Gogoladze:2003ci,Auto:2003ys,Balazs:2003mm,Pallis:2003aw,Profumo:2003ema,Dermisek:2003vn,
Auto:2004km,Gomez:2005nr,Dermisek:2005sw,Baer:2008jn,Altmannshofer:2008vr,Antusch:2008tf,Baer:2008xc,Guadagnoli:2008ui,Baer:2008yd,Antusch:2009gu,Gogoladze:2009ug,
Guadagnoli:2009ze,Baer:2009ie,Enkhbat:2009jt,Baer:2009gg,Baer:2009ff,Choi:2010dk,Gogoladze:2010fu,Gogoladze:2011db,Gogoladze:2011be,Dar:2011sj,Monaco:2011wv, Karagiannakis:2011pb,Gogoladze:2011ce,Ajaib:2011pc,Badziak:2011wm,Badziak:2011he,Hall:1993gn,Hempfling:1993kv,Carena:1994bv,Murayama:1995fn})
and in particular understanding whether Yukawa unification is theoretically consistent and compatible with observation. This is an important and non-trivial question, due in part to the large weak scale threshold corrections (corrections that occur when one matches the SM to the MSSM), especially to the bottom mass\cite{Hall:1993gn, Hempfling:1993kv, Carena:1994bv}. These threshold corrections make unification of third generation Yukawa couplings at the GUT scale dependent on the SUSY spectrum. Thus it is only for certain values of the soft SUSY breaking parameters that Yukawa unification is obtained.

To be more specific, the parameters of the SUSY $SO(10)$ models considered here are
\begin{equation}\label{parameters}
  m_{\nicefrac{1}{2}},~m_{16},~m_{10},~M_D^2,~A_0,~\tan\beta,
  ~\text{
and sign}(\mu ).
\end{equation}
We assume the following boundary conditions at the GUT scale,
\begin{eqnarray}
m_Q^2 = m_E^2 = m_U^2 = m_{16}^2 + M_D^2 \\ \label{3}
m_D^2 = m_L^2 = m_{16}^2 - 3 M_D^2 \\ \label{4}
m_{\nu_R}^2 = m_{16}^2 + 5 M_D^2 \\ \label{5}
m_{H_{u,d}}^2 = m_{10}^2 \mp 2 M_D^2. \label{6}
\end{eqnarray}
Here, the parameters $m_Q^2$, $m_E^2$, $m_U^2$, $m_D^2$, and $m_L^2$ refer to the squared SUSY breaking masses for the left-handed squarks, right-handed sleptons, right-handed up-type squarks, right-handed down-type squarks, and left-handed sleptons respectively. Of course, to obtain physical sfermion masses, one must find the eigenvalues of the relevant sfermion mixing matrix. Similarly $m_{\nu_R}^2$ is the squared mass of the right-handed sneutrino, and $m_{H_{u,d}}^2$ give the squared Higgs mass terms for the Higgs doublets which couple to up-type quarks, and down-type quarks and leptons, respectively. The parameters $m_{\nicefrac{1}{2}}$, $m_{16}$, and $m_{10}$ are the GUT scale masses of the gauginos, the sfermions, and the Higgs soft terms, respectively.  D-term contributions to GUT scale Higgs and sfermion soft terms are parameterized by $M_D^2$. These terms are necessary to ensure radiative electroweak symmetry breaking~\cite{Murayama:1995fn}. While in principle, either sign of $M_D^2$ is possible, we will only consider $M_D^2 > 0$ as this is the sign which lowers $m^2_{H_u}$ with respect to $m^2_{H_d}$ and hence aids in allowing radiative electroweak symmetry breaking when $y_b \sim y_t$. Yukawa unification only occurs for special regions in the parameter space defined above. Generally these are regions with quite heavy scalars
and fairly light gauginos. Typically there is also an inverted mass hierarchy, which is radiatively generated~\cite{Pierce:1996zz, Feng:1998iq, Bagger:1999ty, Bagger:1999sy, Baer:1999md, Baer:2001vw}.

There are several difficulties with this parameter space. First, the relatively light gluino is disfavored by the non-observation of any excess in jets plus missing energy at the LHC
\cite{Aad:2011qa,:2011iu,Aad:2011ib,ATLAS-bjet-lep-missing,ATLAS-jet-missing, 
ATLAS-90,ATLAS-bjet-missing, ATLAS-01lep-missing,Aad:2011xm, Aad:2011ks,
daCosta:2011qk, Aad:2011hh,Chatrchyan:2011zy,CMS-SUS-11,CMS-SUS-10,
CMS-SUS-09,CMS-SUS-05,CMS-SUS-04, Collaboration:2011ida, Chatrchyan:2011ek, 
Chatrchyan:2011bj, Khachatryan:2011tk}.
Secondly, in this region of parameter space, the mostly bino Lightest Supersymmetric Particle (LSP) will tend to have too small of an annihilation cross section to be a (thermal) dark matter candidate~\cite{Baer:2008jn, Baer:2008yd, Auto:2004km}. Finally, raising the sfermion masses tends to suppress contributions to the $g-2$ of the muon, which therefore remain in tension with the experimental value~\cite{Bennett:2006fi}. Thus, we are interested in finding a way to preserve the elegance of Yukawa-unified $SO(10)$ models while also allowing for somewhat heavier gauginos and lighter sfermions.

Our approach will be to unify the top and bottom Yukawa couplings at the GUT scale with the third generation diagonal element of the charged lepton Yukawa matrix, but to introduce an arbitrary mixing angle between the second and third generation charged leptons~\cite{Carena:1999xz}. 
Additionally, we will assume that the right-handed neutrino in the
third generation $\mathbf{16}$ mixes with additional $SU(5)$ 
singlet states at the scale at which $SO(10)$ is broken to $SU(5)$; 
this effectively lowers the neutrino Yukawa coupling.

Assuming that these are the only large mixings, the charged lepton mixing angle will translate to the $\theta_{23}$ of the PMNS matrix~\cite{Pontecorvo:1957cp,Maki:1962mu} which governs the neutrino sector. Thus we can view the parameter which we have added to the standard Yukawa-unified SUSY picture as postdicting (or being constrained by) large mixing in the neutrino sector. This scenario could be embedded into some specific GUT realization. We will not consider a specific embedding here; a discussion of this possibility was presented in~\cite{Carena:1999xz}.

In Section~\ref{scenario} we will describe this alternative scenario, in which second and third generation mixing in the charged lepton sector is used to accommodate third generation Yukawa unification, in more detail. The consequences of this mixing for the neutrino sector will be discussed in Section~\ref{neutrino}. In Section~\ref{scan} we will describe a scan of the parameter space
described in Eq.~(\ref{parameters}).  In particular we will note the preferred values for the mixing angle and its consequences for the neutrino sector in Section~\ref{sin23}, while we will discuss
the collider and dark matter phenomenology of parameter space points allowed by the scan in Section~\ref{pheno}.  We present our conclusions in Section~\ref{conclusions}.

\section{The Scenario}\label{scenario}

In the absence of threshold corrections, Yukawa unification is relatively easy to achieve. In fact values of the top mass close to that which was in fact measured had been predicted in Yukawa unified scenarios (see e.g.~\cite{Ananthanarayan:1991xp, Anderson:1992ba, Hall:1993gn, Carena:1994bv}). Due to the threshold corrections mentioned above, however, the bottom Yukawa at the weak scale is different from what it would be naively. These corrections lower the bottom mass when $\mu M_3 > 0$, as is suggested by the observed discrepancies in $g-2$~\cite{Bennett:2006fi}. In this case, a larger value of $\tan \beta$ than the $\sim 55$ expected from taking the ratio of the running top and bottom masses at the weak scale is required for the unification of $y_t$ and $y_b$ at the GUT scale. The corresponding threshold corrections to the tau mass tend to be smaller than those to the bottom mass, due to the absence of SUSY QCD contributions.  As a result, for the values of $\tan \beta$ required for top-bottom unification with threshold corrections, the tau Yukawa at the GUT scale tends to be larger than the (unified) top and bottom Yukawa couplings.

Here we consider a scenario where instead of unifying the tau Yukawa with the the top and bottom Yukawas at the GUT scale, we take the second and third generations of the charged lepton Yukawa matrix to be described by
\begin{equation}\label{charged lepton 1}
  Y_e(M_{\text{GUT}}) =
  \begin{pmatrix}
    y_{22} & x \\
    x & y_{t/b}
  \end{pmatrix},
\end{equation}
where $y_{t/b}$ is the (unified) top and bottom Yukawa coupling at the GUT scale.  We have limited ourselves to a symmetric matrix for simplicity.  In this paper, we will consider only the second
and third generations.  We do not expect that considering all three generations would qualitatively affect our results, though future research will evaluate this claim.

The form of the Yukawa matrix in Eq.~(\ref{charged lepton 1}) describes a scenario where the third generation Yukawa coupling at the GUT scale is $y_{t/b}$ for an entire $\mathbf{16}$ of $SO(10)$, but the charged leptons in this $\mathbf{16}$ mix with the charged leptons in the second generation $\mathbf{16}$.  To be precise, both the left and right-handed charged leptons in the ``third generation'' $\mathbf{16}$ mix with their counterparts in 
the second generation $\mathbf{16}$ with the same mixing angle $\Theta$.

As a result of the assumption that the mixing angle is the same for both charged lepton states, the transformation of the Yukawa matrix from the basis in which it is diagonal is simply that resulting from a rotation. Thus, due to the invariance of the trace and the determinant under rotations, the parameters in Eq.~(\ref{charged lepton 1}) are given by
\begin{equation}\label{y22}
  y_{22} = y_\tau + y_\mu - y_{t/b}
\end{equation}
and
\begin{equation}\label{x}
  x^2 = (y_{t/b})y_{22} - y_\tau y_\mu.
\end{equation}
In these equations, as in Eq.~(\ref{charged lepton 1}) above, all Yukawas are to be evaluated at the GUT scale.

We can also write the non-diagonal charged lepton Yukawa matrix of Eq.~(\ref{charged lepton 1}) in terms of $y_\tau$, $y_\mu$, and the mixing angle $\Theta$, as follows:
\begin{equation}\label{charged lepton 3}
  \frac{y_\tau + y_\mu}{2}
\begin{pmatrix}
  1 & 0 \\
  0 & 1
\end{pmatrix}
 +
\frac{y_\tau - y_\mu}{2}
\begin{pmatrix}
  - \cos{2\Theta} & \sin{2\Theta} \\
  \sin{2\Theta} & \cos{2\Theta}
\end{pmatrix},
\end{equation}
where $\Theta$ describes the mixing of the charged leptons in the $\mathbf{16}$ with the top and bottom with the charged leptons in another multiplet.
We see immediately that
\begin{equation}\label{sin2 2 Theta}
  \sin^2{2\Theta} = \frac{4 x^2}{(y_\tau - y_\mu)^2},
\end{equation}
where we consider the particular quantity $\sin^2{2\Theta}$ here due to its significance in the induced mixings in the neutrino sector, which are discussed in Section~\ref{neutrino} below.

Using Eq.~(\ref{y22})~and~Eq.~(\ref{x}), Eq.~(\ref{sin2 2 Theta}) can be written explicitly as
\begin{equation}\label{sin2 2 Theta 2}
  \sin^2{2\Theta} = \frac{4(y_{t/b}(y_\tau+y_\mu-y_{t/b}) -y_\tau y_\mu)}
  {(y_\tau - y_\mu)^2}.
\end{equation}
We now define
\begin{equation}\label{R}
  R = \frac{y_\tau}{y_{t/b}}
\end{equation}
for convenience.  We note that in the case where the top and bottom Yukawas are equal at the GUT scale, this corresponds to the $R$ in e.g.~\cite{Baer:2009ie} when $y_\tau > y_{t/b}$, as is in fact obtained in this scenario. Ignoring $y_\mu$, as compared with $y_\tau$, Eq.~(\ref{sin2 2 Theta 2}) can be re-expressed as
\begin{equation}\label{sin2 2 Theta 3}
  \sin^2 2\Theta \approx 4 \frac{R-1}{R^2}
\end{equation}
Thus, we will find $\sin^2 2\Theta \approx 1$ when $R \approx 2$. In fact, this result is quite stable for $R \approx 2$; to be more precise, if $R = 2 + \epsilon$, then Eq.~\ref{sin2 2 Theta 3} becomes
\begin{equation}\label{sin2 2 Theta 4}
  \sin^2 2\Theta \approx \frac{4 + 4\epsilon}{4 + 4 \epsilon + \epsilon^2}
  \approx 1 - \epsilon^2/4.
\end{equation}
Therefore, if we write our charged Yukawa matrix at the GUT scale in the form given in Eq.~(\ref{charged lepton 1}) or Eq.~(\ref{charged lepton 3}),  where the Yukawa matrix has a diagonal element equal to the top and bottom Yukawas at the GUT scale, then the mixing will be maximal when the tau Yukawa at the GUT scale is approximately twice the (unified) top and bottom Yukawa coupling.

The value of $\Theta$ obtained above will, in general, evolve as the charged  lepton Yukawa matrix is run from the GUT or Majorana mass scale to the weak scale. However, in the limit where neutrino Yukawa couplings vanish, every term in the one or two loop $\beta$ functions~\cite{Martin:1993zk} for $\byuk_e$ is either proportional to $\byuk_e$, $\byuk_e \byuk_e^\dagger \byuk_e$, or $\byuk_e \byuk_e^\dagger \byuk_e \byuk_e^\dagger \byuk_e$. Writing $\byuk_e = U^\dagger \mathbf{D}_e U$, we find that we can write $\byuk_e \byuk_e^\dagger \byuk_e = U^\dagger \mathbf{D}_e \mathbf{D}^*_e \mathbf{D}_e U$, etc.  Thus we can write the Renormalization Group Equations (RGE) in the schematic form:
\begin{equation}
  \frac{d ( U^\dagger \mathbf{D}_e U )}{dt}  =
  U^\dagger f(\mathbf{D}_e) U,
\end{equation}
where $f(\bf{D}_e)$ is the right hand side of the RGE in terms of diagonal Yukawa matrices.  Since the RGE (obviously) take the same form for diagonal Yukawa matrices as for general Yukawa matrices, we have that
\begin{equation}
  U^\dagger \frac{d \mathbf{D}_e}{dt} U = U^\dagger f(\mathbf{D}_e) U,
\end{equation}
and, as a result,
\begin{equation}
  \frac{d U^\dagger}{dt} \mathbf{D}_{e} U +
  U^\dagger \mathbf{D}_e \frac{U}{dt} = 0.
\end{equation}
Plugging in the general form of the rotation matrix for $U$ and our specific values for the entries of $\bf{D}_e$, we obtain
\begin{equation}
  \bigg(\frac{d\Theta}{dt}\bigg) (y_\tau - y_\mu)
  \begin{pmatrix}
    \sin{2\Theta} & \cos{2\Theta} \\
    \cos{2\Theta} & -\sin{2\Theta}
  \end{pmatrix} = 0 ,
\end{equation}
which, given the non-degeneracy of $y_\tau$ and $y_\mu$, can only hold if
\begin{equation}
  \frac{d\Theta}{dt} = 0.
\end{equation}
Thus, in the limit where neutrino Yukawa couplings are zero, the value of $\Theta$ (or functions of $\Theta$) obtained, e.g. in Eq.~(\ref{sin2 2 Theta 3}) is the value at any scale, not just at the GUT scale.  While neutrino Yukawa couplings are non-zero, and in fact generally large in Yukawa unified models (at least for the third generation), as mentioned before we will assume a 
dilution of neutrino Yukawa couplings
via the mixing of the right-handed neutrino in the third generation $\mathbf{16}$ with other states.
Therefore the RGE effects on the mixing angle $\Theta$ become negligible.

\section{Neutrino Sector}\label{neutrino}

\subsection{Mixing}

As noted above, in this scenario, the third generation fermions are part of a $\mathbf{16}$ of $SO(10)$ for which Yukawa unification at the GUT scale holds, but where the right- and left-handed charged leptons in this multiplet mix with the right- and left-handed charged leptons in the second generation $\mathbf{16}$. We assume that there is no similar inter-generation mixing in the neutrino sector.

As the $W$ will only couple states within these multiplets, the left-handed neutrino in the $\mathbf{16}$ will couple to a linear combination of the physical muon and tau. Considering instead the combinations of left-handed neutrino eigenstates from the two multiplets which couple to a given charged lepton mass eigenstate, we find
\begin{equation}\label{nu mu}
  \nu_\mu= \cos{\Theta}~\nu_2 - \sin{\Theta}~\nu_3 ,
\end{equation}
and
\begin{equation}\label{nu tau}
  \nu_\tau= \sin{\Theta}~\nu_2 + \cos{\Theta}~\nu_3,
\end{equation}
where we have defined the non-diagonal charged lepton matrix in terms of the mixing angle $\Theta$, as in Eq.~(\ref{charged lepton 3}) above. This gives the $2$ by $2$ PMNS matrix $U$ defined by
\begin{equation}\label{PMNS}
  \begin{pmatrix}
    \nu_\mu  \\
    \nu_\tau
  \end{pmatrix} =
  \begin{pmatrix}
    \cos{\Theta} & - \sin{\Theta} \\
    \sin{\Theta} & \cos{\Theta}
  \end{pmatrix}
  \begin{pmatrix}
    \nu_2  \\
    \nu_3
  \end{pmatrix}
\end{equation}
From this we find that
\begin{equation}
  \theta_{23} = -\Theta
\end{equation}
Thus, in particular, we find
\begin{equation}\label{neutrino mixing conclusion}
  \sin^2 2 \theta_{23} = \sin^2 2 \Theta.
\end{equation}

\subsection{Yukawa Couplings}

It has been noted, for example in~\cite{Borzumati:1986qx, Hisano:1995nq,Masiero:2002jn,Campbell:2003wp,Masiero:2004vk, Masiero:2004js}, that large values of neutrino Yukawa couplings generically lead to large lepton flavor violation (though for a somewhat contrary perspective see~\cite{Barger:2009gc}). This flavor violation arises from the contribution of this Yukawa term to the RGE running of the slepton mass matrix.  As noted above, we assume a dilution of neutrino Yukawa couplings, e.g. through additional mixing between the right-handed neutrinos and unspecified additional states at the GUT scale.  
We demand sufficient dilution to reduce $BR (\mu \to e \gamma)$ 
to below the experimental bounds, in particular the $90$\% 
C.L. upper limit on this quantity, $2.4 \times 10^{-12}$~\cite{Adam:2011ch}).
We will examine this mixing and the resultant dilution of the neutrino 
Yukawa couplings in greater detail in future work.

\section{Scan of Parameter Space}\label{scan}

As a step toward determining which regions in the parameter space of this scenario are allowed by existing constraints, we performed a scan of the parameter space using the SUSY spectrum generation and analysis code \texttt{SOFTSUSY~3.1.7}~\cite{Allanach:2001kg}, as well as the dark matter calculation code \texttt{DarkSUSY~5.0.5},~\cite{Gondolo:2004sc} and the code for NLO calculation of $BR( b \to s \gamma)$ in the MSSM, \texttt{SusyBSG~1.4}~\cite{Degrassi:2007kj}. We modified the GUT-scale boundary conditions in \texttt{SOFTSUSY} to be those of this scenario (Eq.~(\ref{3}) - (\ref{6})).

In this scan, the supersymmetric parameter $\mu$ was always chosen to be positive, in order to reduce the tension with the observed value of muon $g-2$~\cite{Bennett:2006fi}. The ranges for the other parameter regions in which the points are found are given in Table~\ref{parameter}; points were scanned using a grid search. 
We found that with the exception of the upper bound for $m_{16}$, the
specific values of the bounds used in the grid search were not 
particularly important, as very few points in the parameter space that
satisfied the constraints we impose (detailed below) were near the 
boundary of the parameter space scanned.  The upper bound on $m_{16}$
was used to restrict our focus to regions of parameter space with 
somewhat lighter sfermions.
\begin{table}\label{parameter}
  \begin{center}
    \begin{tabular}{|l|c|c|c|}
      \hline
      Parameter & Min & Max & Step width\\
      \hline
      \hline
      $m_{\nicefrac{1}{2}}$ & 400 GeV & 2000 GeV & 25 GeV \\
      \hline
      $m_{16}$ & 300 GeV & 2000 GeV & 20 GeV \\
      \hline
      $\nicefrac{m_{10}}{m_{16}}$ & 1.1 & 1.4 & 0.05 \\
      \hline
      $\nicefrac{M_D}{m_{16}}$ & 0.275 & 0.4 & 0.025\\
      \hline
      $A_0$ & -2000 GeV & -200 GeV & 30 GeV\\
      \hline
      $\tan{\beta}$ & 58 & 64 & 0.3 \\
      \hline
      \hline
    \end{tabular}
    \caption{Ranges of parameters used in the parameter
    scan described in Section 4.}
  \end{center}
\end{table}

A given point was only considered for further study if the following criteria were satisfied
\begin{enumerate}
\item \textbf{Consistency of spectrum.}
Specifically none of the ''problem flags'' listed in Appendix B of Ref.~\cite{Allanach:2001kg} are triggered. These flags are triggered if the spectrum calculation does not converge to the specified tolerance, if couplings become non-perturbative, if a Landau pole is encountered in the RGE running, if electroweak symmetry is not broken, if there are tachyons in the model,
if the $\rho$ parameter cannot be calculated, if the Higgs potential is unbounded from below, or if the GUT scale is less than $10^4$ or greater than $5 \times 10^{17}$ GeV.
\item \textbf{Top/ bottom Unification.}
The top and bottom have the same Yukawa coupling at the GUT scale.  Specifically we demand
\begin{equation}
| \frac{y_t - y_b}{y_t + y_b} | < 0.02
\end{equation}
where $y_t$ and $y_b$ are the top and bottom Yukawa couplings at the GUT scale.
\item \textbf{$R$ in desired range.}
In the scenario described, if $R > 2$, then $y_{22} > y_{33}$. ($R$ is defined in Eq.~(\ref{R}).)  We have restricted our analysis to the case in which $y_{33} > y_{22}$ and hence $R<2$.  Very few parameter points were eliminated by this constraint.
\item \textbf{SUSY spectrum constraints.}
Charged sparticles must be heavy enough to have evaded detection at LEP; specifically we demand that the chargino mass be greater than $94$ GeV. Correspondingly, the neutralino mass is greater than $46$ GeV. In addition, we require that the light SM-like Higgs mass is greater than $114$ GeV. We note that some of these constraints on the SUSY spectrum are conservative; our main concern is with determining the general properties of the allowed parameter space, rather than exploring ways in which naive bounds could be circumvented.
\item \textbf{Dark Matter Constraints.}
We demand that the LSP be a neutralino, and that its thermal relic density satisfy $0.1<\Omega_\chi h^2<0.12$, following~\cite{Hinshaw:2008kr}. 
We implement the bound on the spin-independent dark matter direct
detection cross section from XENON100~\cite{Aprile:2011hi}, 
which is $\sim 10^{-44}$~cm$^2$ for most LSP masses.
Likewise, to avoid bounds e.g. from COUPP\cite{Behnke:2010xt}, we demand that the spin-dependent WIMP-nucleon cross section be less than $10^{-41}$~cm$^2$.
\item \textbf{B Physics Constraints.}
We demand that the calculated value of $BR( b \to s \gamma)$ 
be in a range consistent with the world-average experimental value 
$3.55 \pm 0.24 \pm 0.09$~\cite{Asner:2010qj}. 
Specifically, we demand that the value of $BR( b \to s \gamma)$, 
calculated with \texttt{SusyBSG 1.4}~\cite{Degrassi:2007kj}, 
be in the range 
$3.04 \times 10^{-4} < BR( b \to s \gamma) < 4.06 \times 10^{-4}$.
Likewise, we demand $BR( B_s \to \mu^+ \mu^-) < 1.2 \times 10^{-8}$,
for consistency with recent LHCb and CMS searches~\cite{Bettler:2011rp,
Serrano:2011px,Chatrchyan:2011kr,CMS_plus_LHCb,Akeroyd:2011kd} 
for this rare decay.
We do not add a separate contraint from $BR(B \to \tau \nu)$, 
as the value of this quantity in the parameter space points which 
satisfy the preceding constraints, epsecially the limit on
$BR( B_s \to \mu^+ \mu^- )$, is always consistent with experiment
as will be discussed in more detail below.
Both $BR(B_s \to \mu^+ \mu^-)$ and $BR(B \to \tau \nu)$ were calculated 
with \texttt{SuperIso 3.2}~\cite{Mahmoudi:2008tp,Mahmoudi:2009zz};
the calculations for all Higgs masses and B physics observables
were verified using \texttt{CPsuperH2.0}~\cite{Lee:2007gn}.

\end{enumerate}

\section{Values of $\sin^2 2\theta_{23}$}\label{sin23}

\begin{figure}
\centering
\includegraphics[height=2.8in]{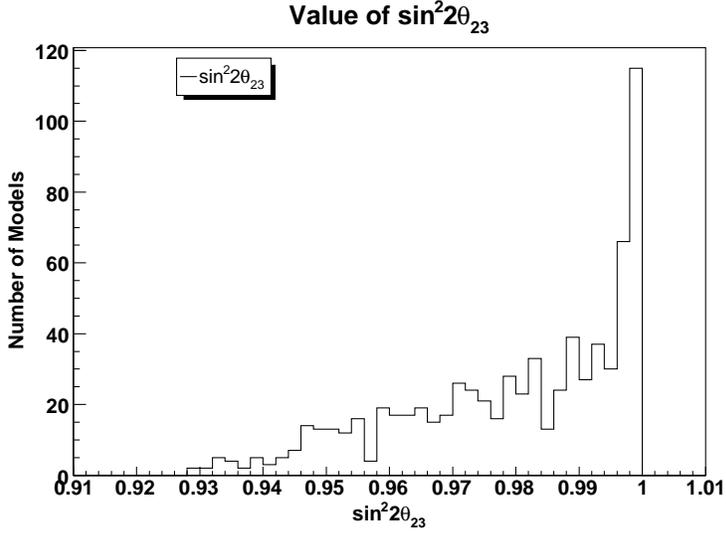}
\caption{Distribution of $\sin^2{\theta_{23}}$ (obtained from taking $\theta_{23} = - \Theta$, and using Eq.~(\ref{sin2 2 Theta 2}) to determine $\Theta$) for the models found in the scan. }
\label{sin2 2 theta23}
\end{figure}

The distribution of $\sin^2 2\theta_{23}$ for models obtained in the scan described above is presented in Fig.~\ref{sin2 2 theta23}. Note that all of the models found in the scan are consistent with the experimental observation that $\sin^2 2\theta_{23} \approx 1$. We now examine why this relation holds so generally.

\begin{figure}
\centering
\includegraphics[height=2.8in]{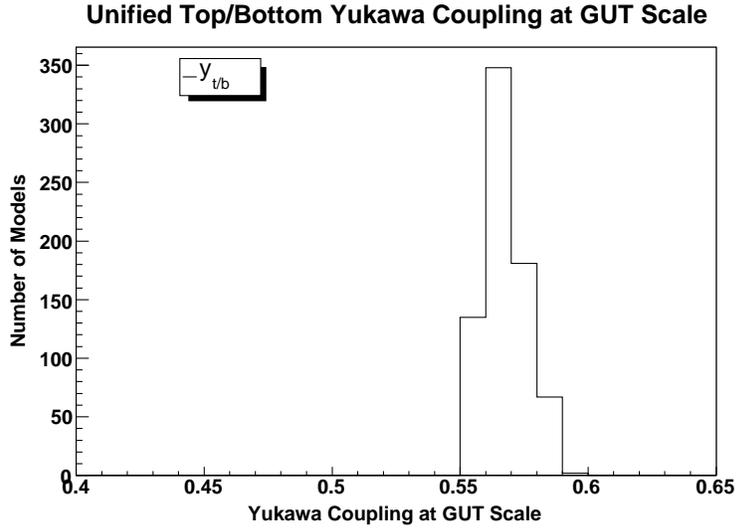}
\caption{Distribution of the (unified) top/bottom Yukawa coupling at the GUT scale in the models found from the scan. (In our scan, we demand that these top Yukawa couplings be unified to within $4\%$ at the GUT scale; we take the geometric mean of these quantities when constructing these histograms.)}
\label{ytb}
\end{figure}

\begin{figure}
\centering
\includegraphics[height=2.8in]{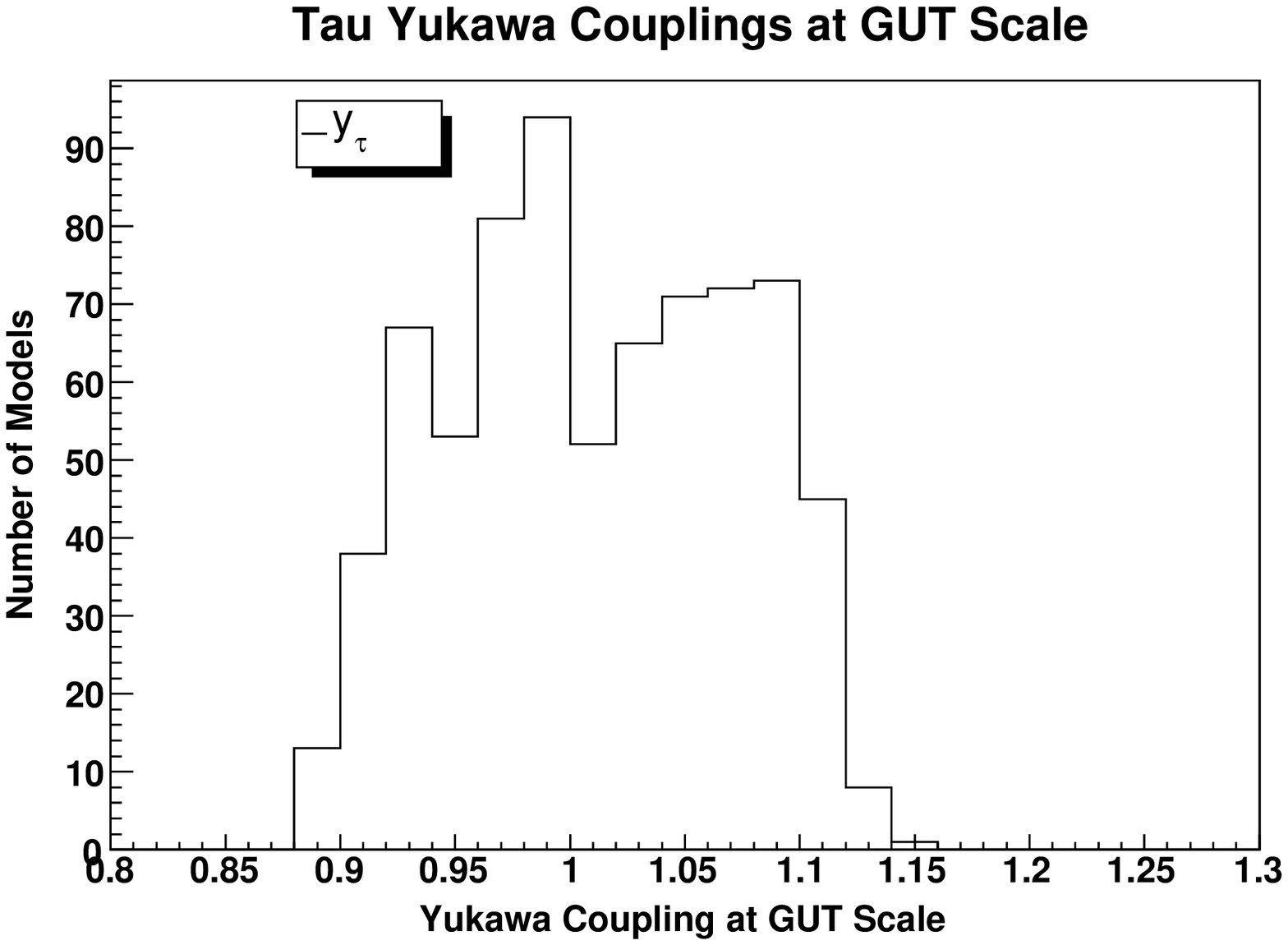}
\caption{Distribution of the tau Yukawa coupling at the GUT scale in the models found from the scan.}
\label{ytau}
\end{figure}

The value of $\sin^2 2\theta_{23}$ is found from Eq.~(\ref{sin2 2 Theta 2}). As we see from the approximation in Eq.~(\ref{sin2 2 Theta 3}), this value is mainly a function of the (unified) top and bottom Yukawa coupling $y_{t/b}$ as well as the tau Yukawa coupling $y_\tau$, defined at the GUT scale. The distribution of $y_{t/b}$ and $y_\tau$ are shown in Figs.~\ref{ytb}~and~\ref{ytau} respectively.  We note that the unified top/bottom Yukawa coupling is always 
in the range $0.55 - 0.59$, while the tau Yukawa coupling ranges from $\sim 0.9$ to $\sim 1.1$.  We note, following Eq.~(\ref{sin2 2 Theta 3}), that the lower end of this range corresponds to values of $\sin^2 2\theta_{23}$ of $\sim 0.93$, while the upper end of the range corresponds to values of $\sim 1$.  Thus having $\sin^2 2 \theta_{23}$ in the desired range is a result of the relatively large tau Yukawa coupling.

We can understand the large Yukawa coupling as follows. In the absence of SUSY corrections to the bottom mass, the value of $\tan{\beta}$ which leads to top-bottom unification yields a value for the GUT scale tau Yukawa coupling which is similar to the value for the 
top and bottom Yukawa couplings. Standard Yukawa unification is not hard to obtain in this scenario. Adding the large negative contributions to the bottom mass raises the value of $\tan{\beta}$ needed to obtain top-bottom unification. This has relatively little effect on the value of the unified top-bottom Yukawa coupling, but serves to raise the tau Yukawa coupling into the desired range, as the tau Yukawa is also enhanced by $\tan{\beta}$.

\begin{figure}
\centering
\includegraphics[height=2.8in]{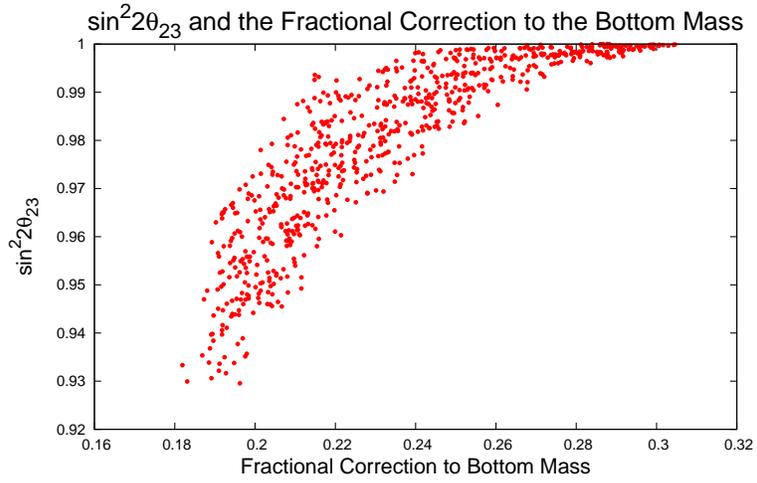}
\caption{Distribution of threshold correction to the bottom mass as a fraction of the bottom mass and the value of $\sin^2{2 \theta_{23}}$ (obtained as above) in models found in the scan.}
\label{deltab_sin2}
\end{figure}

\begin{figure}
\centering
\includegraphics[height=2.8in]{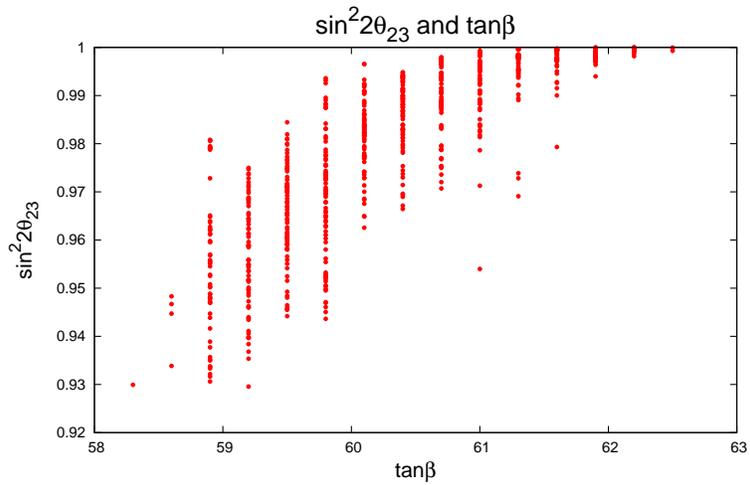}
\caption{Distribution of the values of $\tan{\beta}$ and $\sin^2{2 \theta_{23}}$ (obtained as above) in models found in the scan.}
\label{tanb_sin2}
\end{figure}

Hence we would expect the (absolute value) of the fractional correction to the bottom mass to correlate with larger values of $\sin^2{2\theta_{23}}$, which in fact we do find (as shown in Fig.~\ref{deltab_sin2}).  Additionally, we would expect that the value of $\tan{\beta}$ is correlated with $\sin^2{2\theta_{23}}$. This expectation is confirmed in Fig.~\ref{tanb_sin2}. Since the dominant corrections to the bottom mass are proportional to the gluino mass, but inversely proportional to the square of the sbottom mass, we expect that models with heavier gauginos, i.e. larger values of $m_{\nicefrac{1}{2}}$ will have larger corrections to the bottom mass, and hence larger values of $\sin^2{2\theta_{23}}$. On the other hand, we expect larger values of $m_{16}$, which
correspond to larger values of the sbottom mass, to suppress the SUSY correction to the bottom mass and hence the value of $\sin^2{\theta_{23}}$ obtained. These expectations for the dependence of $\sin^2{2\theta_{23}}$ on $m_{\nicefrac{1}{2}}$ and $m_{16}$ are born out in the models found by the scan, though the dependence in each case is somewhat weak.

\section{Phenomenology}\label{pheno}

We now consider the consequences of this scenario for collider and dark matter phenomenology. In each case, an important difference between the scenario described here and more standard Yukawa-Unified SUSY scenarios, e.g. as considered in ~\cite{Baer:2009ff}, is that third generation sfermions may be relatively light in this scenario. The consequences of this difference, as well as
the general signatures to be expected from points in the parameter space found by our scan (which we will refer to as ``models'' for convenience), will be discussed below.

\subsection{Higgs Phenomenology}

\begin{figure}
\centering
\includegraphics[height=2.8in]{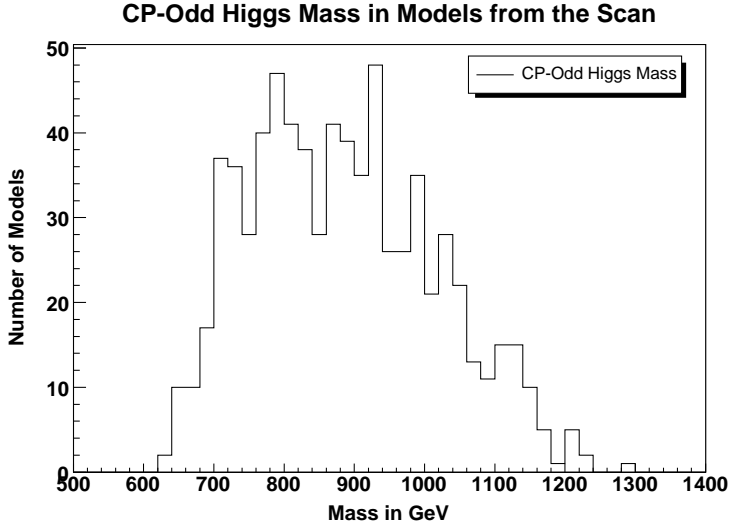}
\caption{The distribution of the CP-odd Higgs mass is shown in this histogram.  The heavy CP-even Higgs and the charged Higgses are nearly degenerate with the CP-odd Higgs in every model found by the scan, so this histogram also describes the mass distribution of these other Higgses.}
\label{cp-odd higgs}
\end{figure}

\begin{figure}
\centering
\includegraphics[height=2.8in]{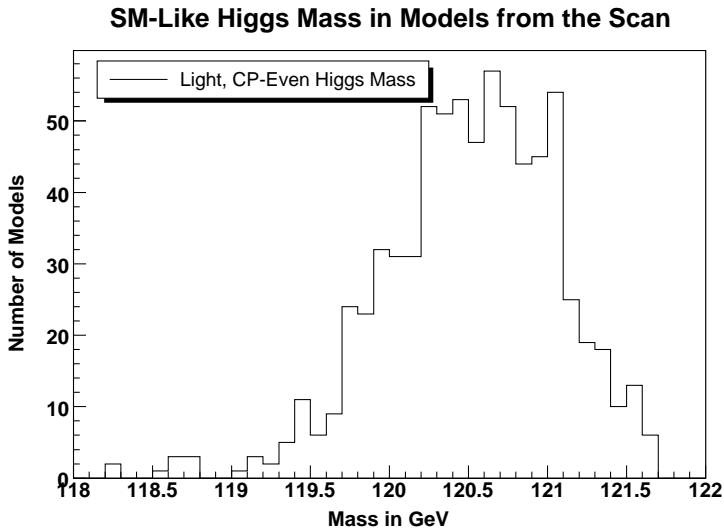}
\caption{This histogram shows the distribution of light, CP-even Higgs masses in models found by the scan.}
\label{higgs}
\end{figure}

Models found by the scan all have decoupled Higgs sectors; i.e., in all models the CP-odd Higgs is heavy and the heavy CP-even Higgs and the charged Higgs bosons are nearly degenerate with it.
(Specifically, $0.993~m_A < m_H < m_A$ and $1.001~m_A < m_{H^\pm} < 1.010~m_A$ in all models found by the scan.)  The distribution of CP-odd Higgs masses is shown in Fig.~\ref{cp-odd higgs}. In every model found by the scan, the light CP-even (Standard Model-like) Higgs has a mass between $118$ and $122$ GeV; the distribution of these masses in the model set is shown in Fig.~\ref{higgs}. 
Thus a Higgs mass between $118$ and $122$ GeV is a prediction of this scenario, up to uncertainties in the Higgs mass calculation.

\subsection{Flavor Physics and Precision Observables}

\begin{figure}
\centering
\includegraphics[height=2.8in]{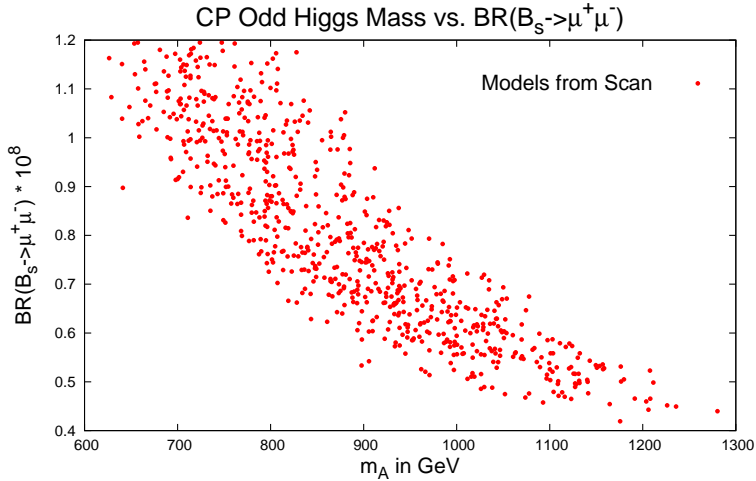}
\caption{This figure shows the dependence of $BR(B_s \to \mu^+ \mu^-)$ 
on the charged Higgs mass.}
\label{bsmm}
\end{figure}

One can observe from Fig.~\ref{cp-odd higgs} that the 
CP-odd Higgs tends to be somewhat heavy.
This is due in part to the constraint on $BR(B_s \to \mu^+ \mu^-)$,
which is effectively a constraint on the CP-odd and charged Higgs
masses. 
As was noted above, the heavy CP-even Higgs and the charged 
Higgs bosons are nearly degenerate with the CP-odd Higgs 
in all of the models found by the scan.
The dependence of the values of $BR(B_s \to \mu^+ \mu^-)$ 
on the CP-odd Higgs mass is shown in Fig.~\ref{bsmm}.
While other parameters beside the CP-odd Higgs mass affect 
the value obtained for this rate, the requirement that
$BR(B_s \to \mu^+ \mu^-) < 1.2 \times 10^{-8}$ 
effectively sets a lower limit of $\sim 600$ GeV on the CP-odd Higgs mass.

\begin{figure}
\centering
\includegraphics[height=2.8in]{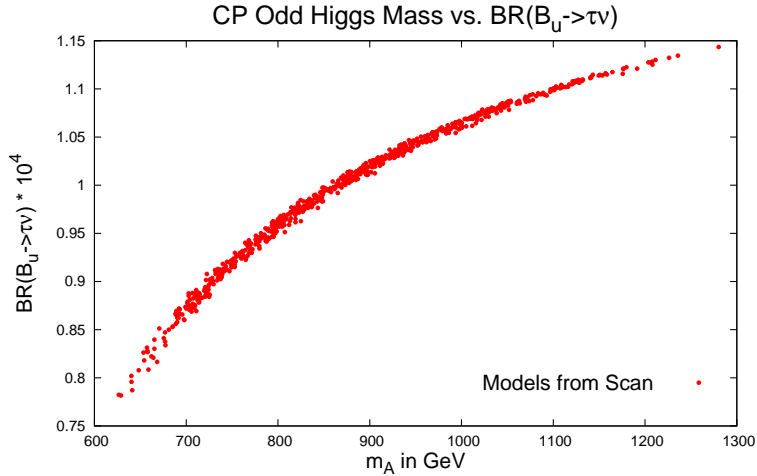}
\caption{This figure shows the dependence of $BR(B \to \tau \nu)$ 
on the charged Higgs mass.}
\label{btnh}
\end{figure}

Fig.~\ref{btnh} shows the dependence of $BR(B \to \tau \nu)$
on the CP-odd Higgs mass (which is nearly degenerate with the
charged Higgs mass).
The values of $BR(B \to \tau \nu)$ shown in this figure
were calculated using 
\texttt{SuperIso 3.2}~\cite{Mahmoudi:2008tp,Mahmoudi:2009zz}, 
but have been scaled so that the SM value is
$1.22 \pm 0.31 \times 10^{-4}$, which is calculated
using the value of $V_{ub}$ obtained from inclusive
decays~\cite{Lunghi:2010gv}.
Using the experimental average for this branching ratio from the HFAG 
collaboration~\cite{Asner:2010qj,HFAGwebpage},
$1.64 \pm 0.34 \times 10^{-4}$, 
we find that the $2 \sigma$ lower bound on the ratio of
the MSSM value for $BR( B \to \tau \nu )$ to the SM value is 
$0.46$.
This corresponds to a value for $BR( B \to \tau \nu)$ of 
$0.56 \times 10^{-4}$, which is somewhat below the
minimum value obtained for parameter space points found by our scan.  
Due to the $1/m_{H^\pm}^2$ dependence of the supersymmetric
contribution to the decay amplitude, the smallest values of this decay
rate are obtained for the smallest values of the CP-odd mass, as it is
clearly seen in Fig.~\ref{btnh}.  In this case,
$BR ( B \to \tau \nu )$ is large enough when the 
CP-odd Higgs mass is $\gtrsim 600$ GeV,  
which was necessary for the $BR ( B_s \to \mu^+ \mu^- )$
constraint to be fulfilled.

We note that there is tension between the SM theory value and the
experimental average for $BR ( B \to \tau \nu )$, and that this tension
is somewhat exacerbated in the parameter space points found by
our scan.  This is in fact a generic feature of SUSY models.
This tension is significantly increased if one uses the value of $V_{ub}$
obtained from exclusive decays; in this case the SM prediction is
$0.67 \pm 0.15 \times 10^{-4}$~\cite{Lunghi:2010gv},
and the $2\sigma$ lower limit on the ratio of
MSSM $BR ( B \to \tau \nu )$ to SM $BR ( B \to \tau \nu )$
is $0.95$.  This corresponds to a value of $0.64 \times 10^{-4}$.
However, if $0.67 \pm 0.15 \times 10^{-4}$ is used for the SM theory value, 
the values displayed in Fig~\ref{btnh} would need to be scaled down by 
$1.22/0.67 = 1.8$ (as it is the ratio of MSSM to SM rate which has been 
calculated).  
Thus this constraint would in fact rule out all points found by our scan,
effectively demanding a CP-odd Higgs mass $\gtrsim 1300$ GeV.

\begin{figure}
\centering
\includegraphics[height=2.8in]{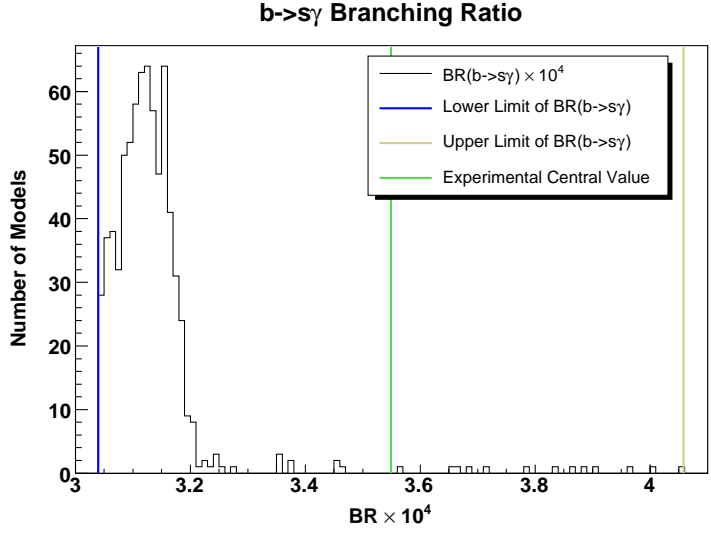}
\caption{This histogram shows the distribution of $BR( b \to s \gamma)$ in the models found by the scan. The upper and lower limits for $BR( b \to s \gamma)$ which we impose are also shown.}
\label{bsg}
\end{figure}

\begin{figure}
\centering
\includegraphics[height=2.8in]{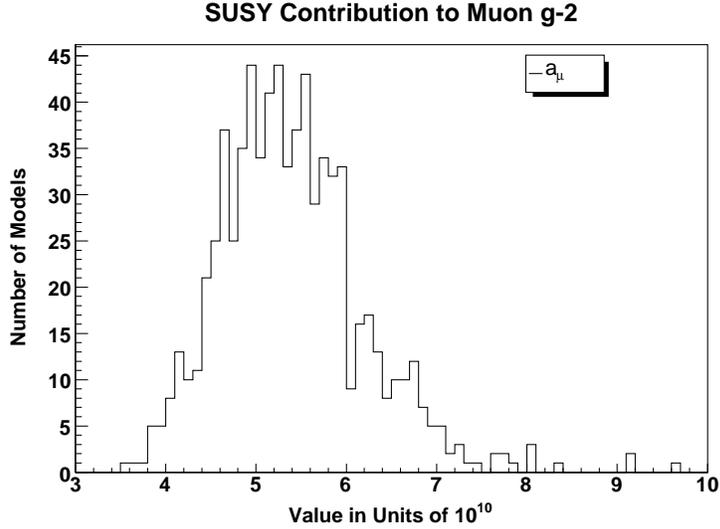}
\caption{This histogram shows the distribution of SUSY contribution to the anomalous magnetic moment of the muon ($a_\mu = (g-2)/2$) in models found by the scan. The one-loop calculation was done
  with \texttt{DarkSUSY}; the two-loop contribution was calculated with \texttt{SoftSUSY}.}
\label{g-2}
\end{figure}

The distribution of the branching ratio for $b \to s \gamma$ is shown in Fig.~\ref{bsg}. We note that the models found by the scan predominantly have values of this quantity, which like the SM value
($3.15 \pm 0.23$~\cite{Misiak:2006zs} ) are somewhat below the world-average experimental value~\cite{Asner:2010qj}, though it should be noted that the theory error is comparable to this discrepancy.
The SUSY contribution to the anomalous magnetic moment of the muon is shown in Fig.~\ref{g-2}. We note that the value of this quantity obtained in these models generally reduces the tension between the SM value and experiment (the experimental value is higher by $(22.4 \pm 10) \times 10^{-10}$ to $(26.1 \pm 9.4) \times 10^{-10}$ ~\cite{Bennett:2006fi} depending on the precise method used
to calculate the SM contribution). However the size of the SUSY contribution in models found by the scan is insufficient to account for the experimental central value; generic models found by the scan have a SUSY contribution of $\sim (4-8) \times 10^{-10}$. An extension of scenario presented here, in which the first and second generation sfermion mass terms do not have the same boundary conditions as the third generation sfermion mass terms, might allow for $g-2$ values closer to the observed value.

\subsection{Gluino Production and Cascade Decays}

The gluino mass and the decay modes of the gluino play a large role in determining what signatures a given SUSY scenario 
will have at the LHC. Models found by our scan have gluinos in the $\sim 1.5-3$ TeV range. This is partially a result 
of the range scanned, however as $m_{\nicefrac{1}{2}}$ was scanned up to $2$ TeV, the absence of models allowed by 
our scan with gluino masses $\gtrsim 3$ TeV, suggests that models with large values of $m_{\nicefrac{1}{2}}$ 
are somewhat more likely to run afoul of our constraints, in particular the constraint that the thermal 
relic density be consistent with the WMAP value.

\begin{figure}
\centering
\includegraphics[height=2.8in]{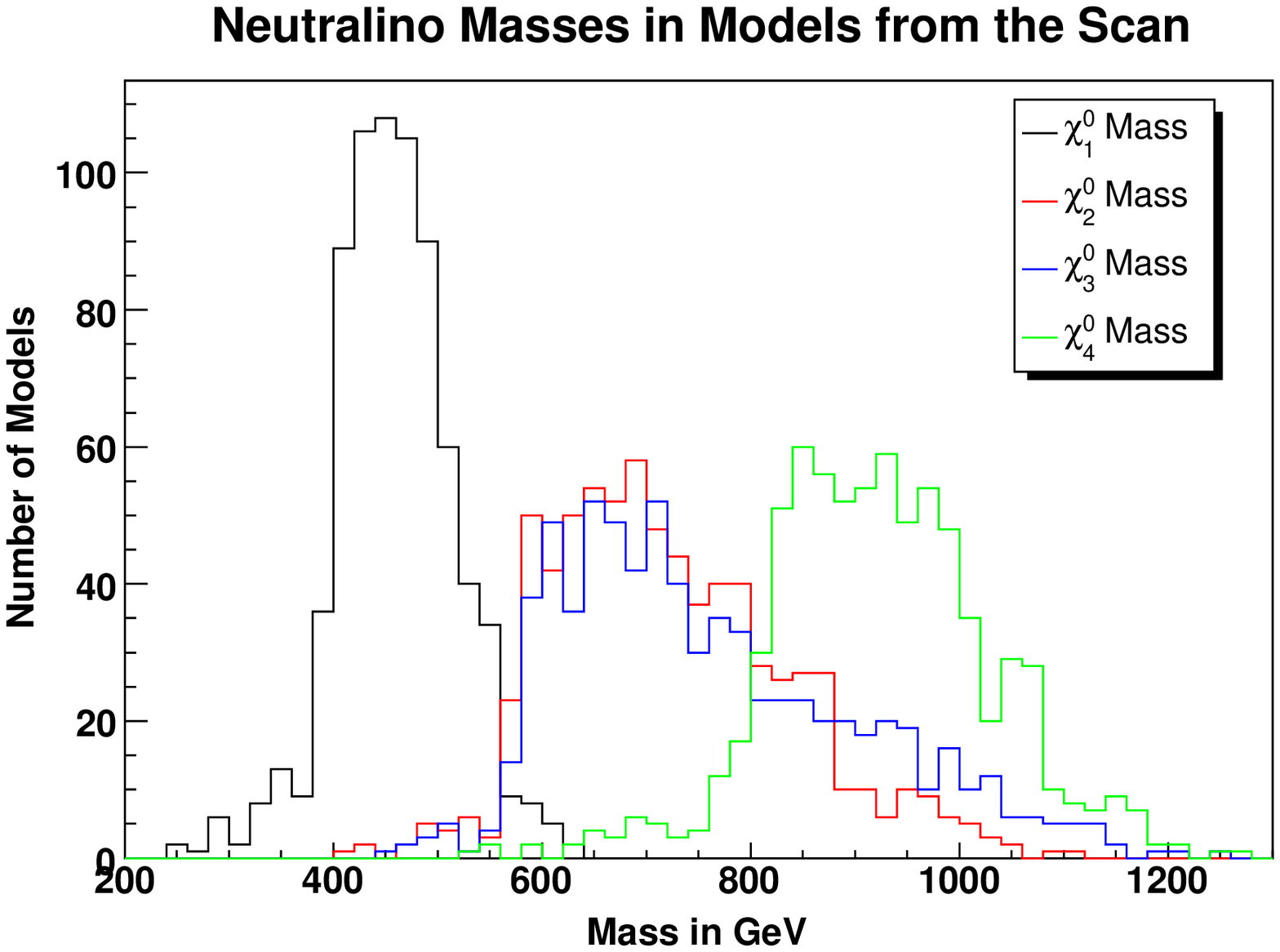}
\caption{This histogram shows the distribution of neutralino masses in models found by the scan.}
\label{neutralino}
\end{figure}

\begin{figure}
\centering
\includegraphics[height=2.8in]{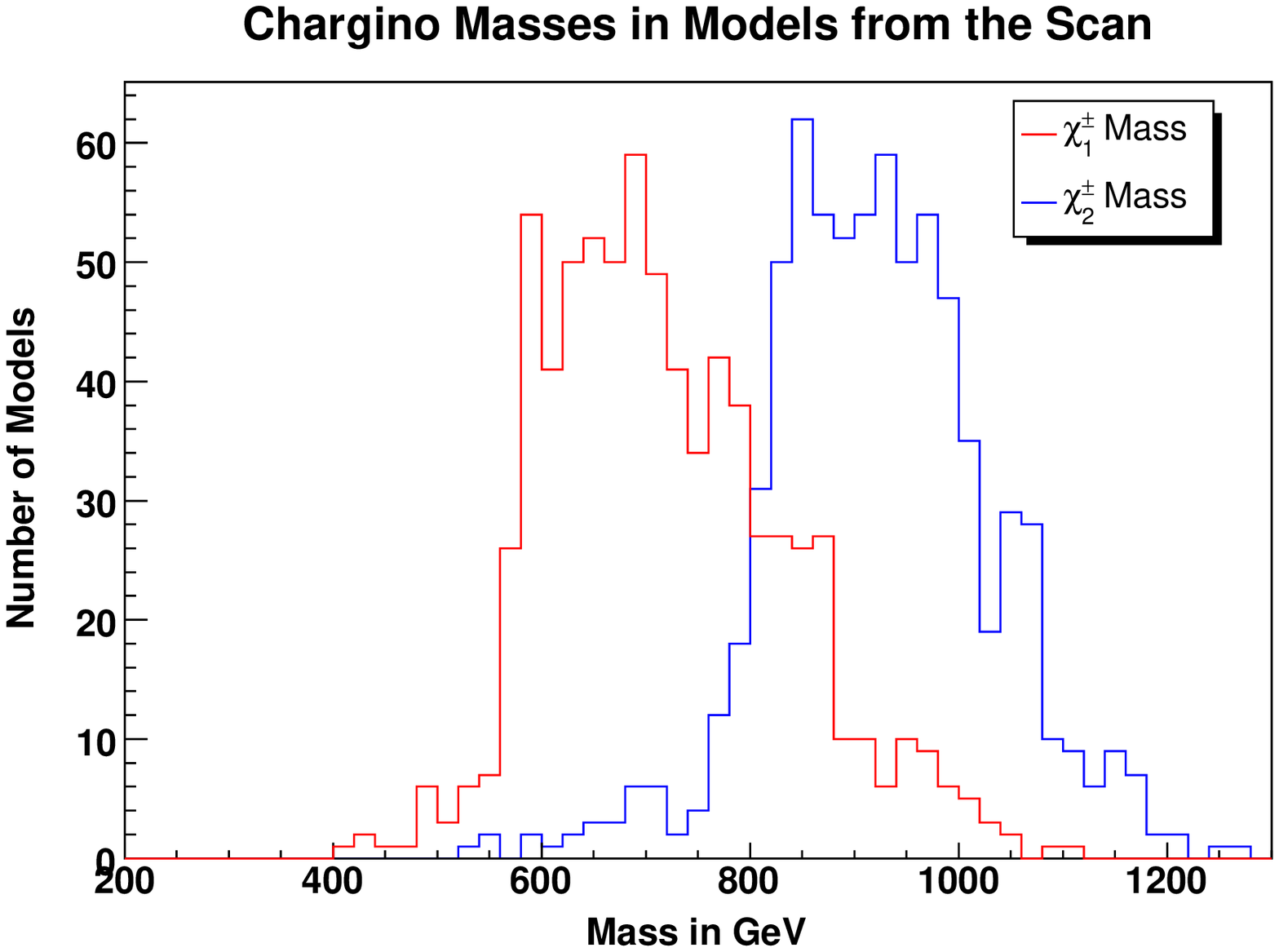}
\caption{This histogram shows the distribution of chargino masses in models found by the scan.}
\label{chargino}
\end{figure}

\begin{figure}
\centering
\includegraphics[height=2.8in]{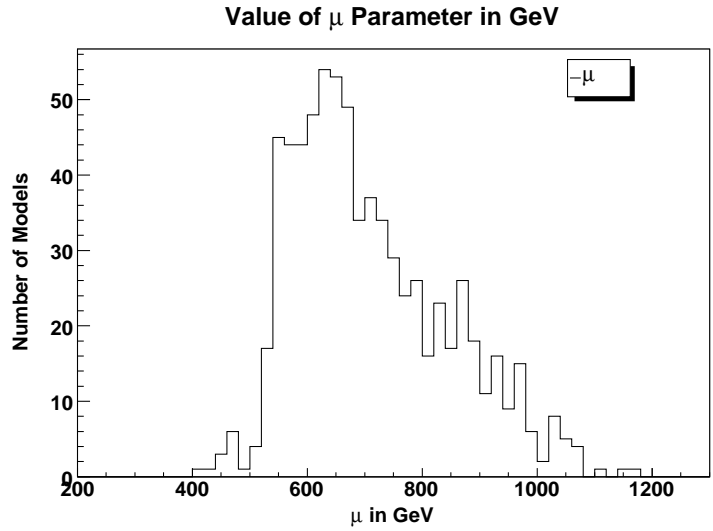}
\caption{This histogram shows the distribution of the Higgsino mass parameter $\mu$ in models found by the scan.}
\label{mu}
\end{figure}

\begin{figure}
\centering
\includegraphics[height=2.8in]{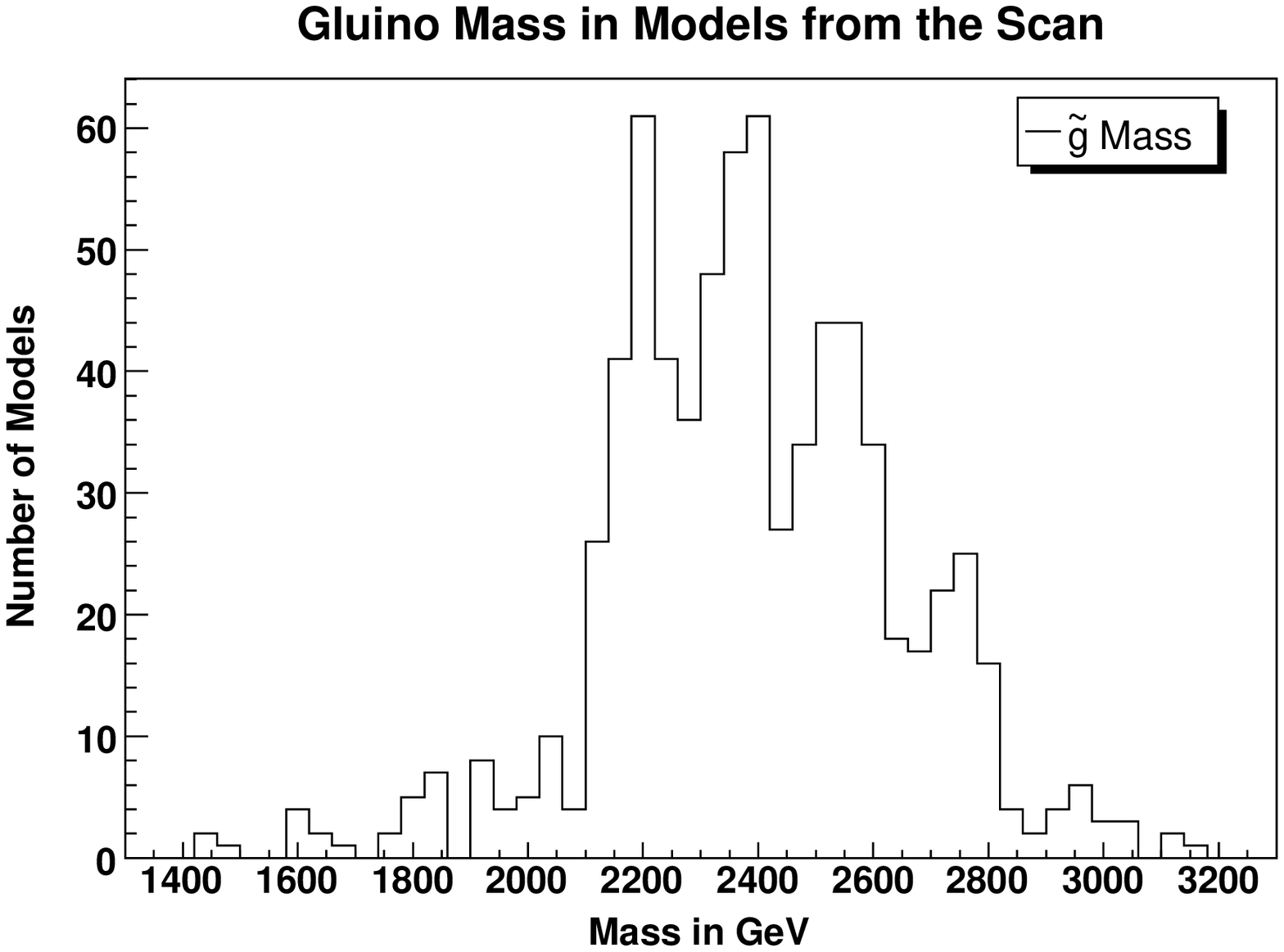}
\caption{This histogram shows the distribution of gluino masses in models found by the scan.}
\label{gluino}
\end{figure}

\begin{figure}
\centering
\includegraphics[height=2.8in]{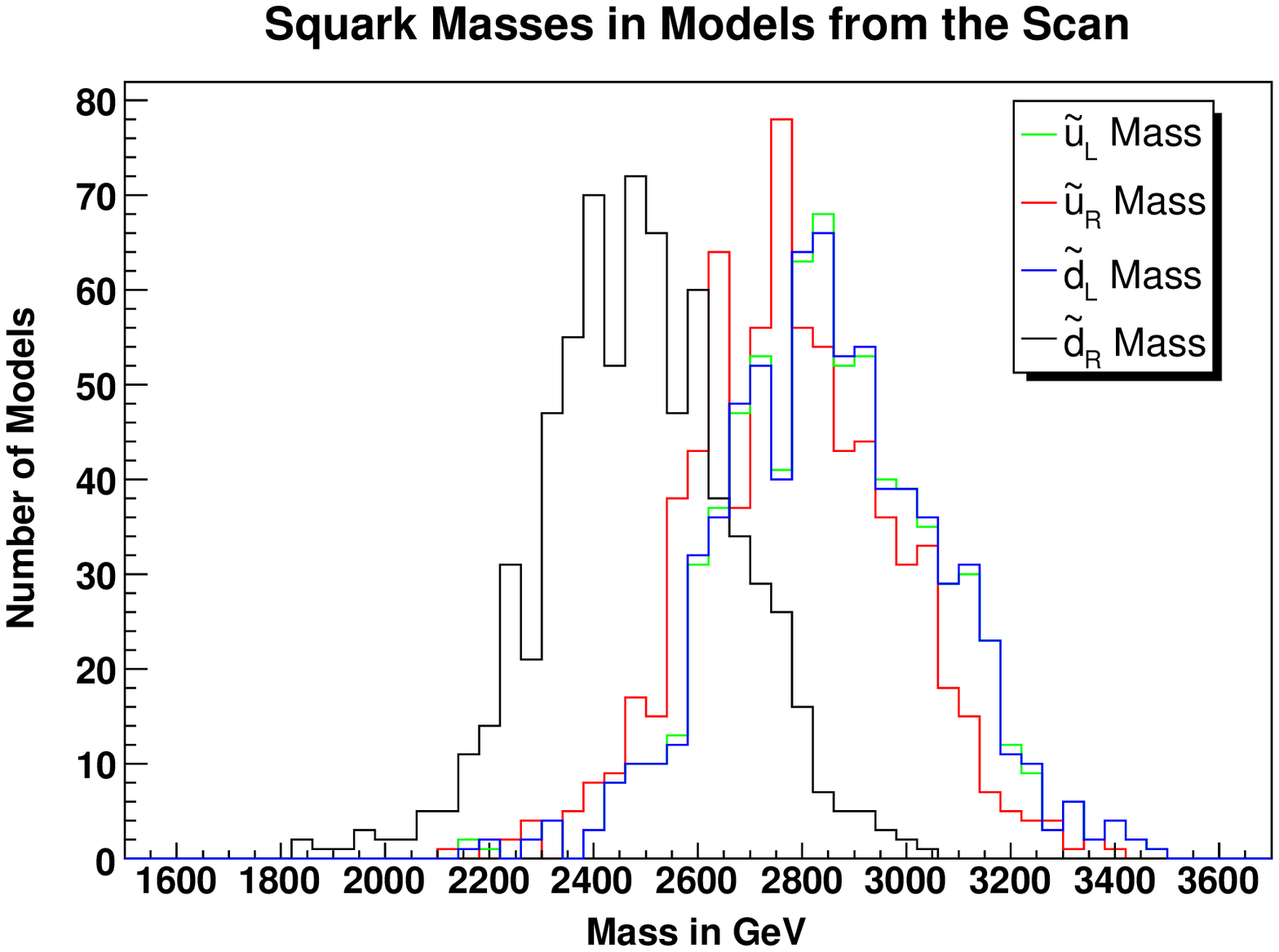}
\caption{This histogram shows the distribution of the first and second generation squark masses in models found by the scan.}
\label{squarks}
\end{figure}

Gaugino mass unification at the GUT scale is an assumption of this scenario; this means that there is a mostly bino 
LSP with mass in the $\sim 250 - 600$ GeV range, and a mostly wino chargino and neutralino with masses $\sim 2m_{LSP}$. 
The Higgsino-like charginos and neutralinos generally have masses similar to the wino-like ones.  
The distribution of neutralino masses in these models is shown in Fig.~\ref{neutralino}, 
while Fig.~\ref{chargino} shows the distribution of chargino masses; Fig.~\ref{mu} shows the distribution 
of the $\mu$ parameter in models found by the scan. The distribution of gluino masses is included in Fig.~\ref{gluino}.

\begin{figure}
\centering
\includegraphics[height=2.8in]{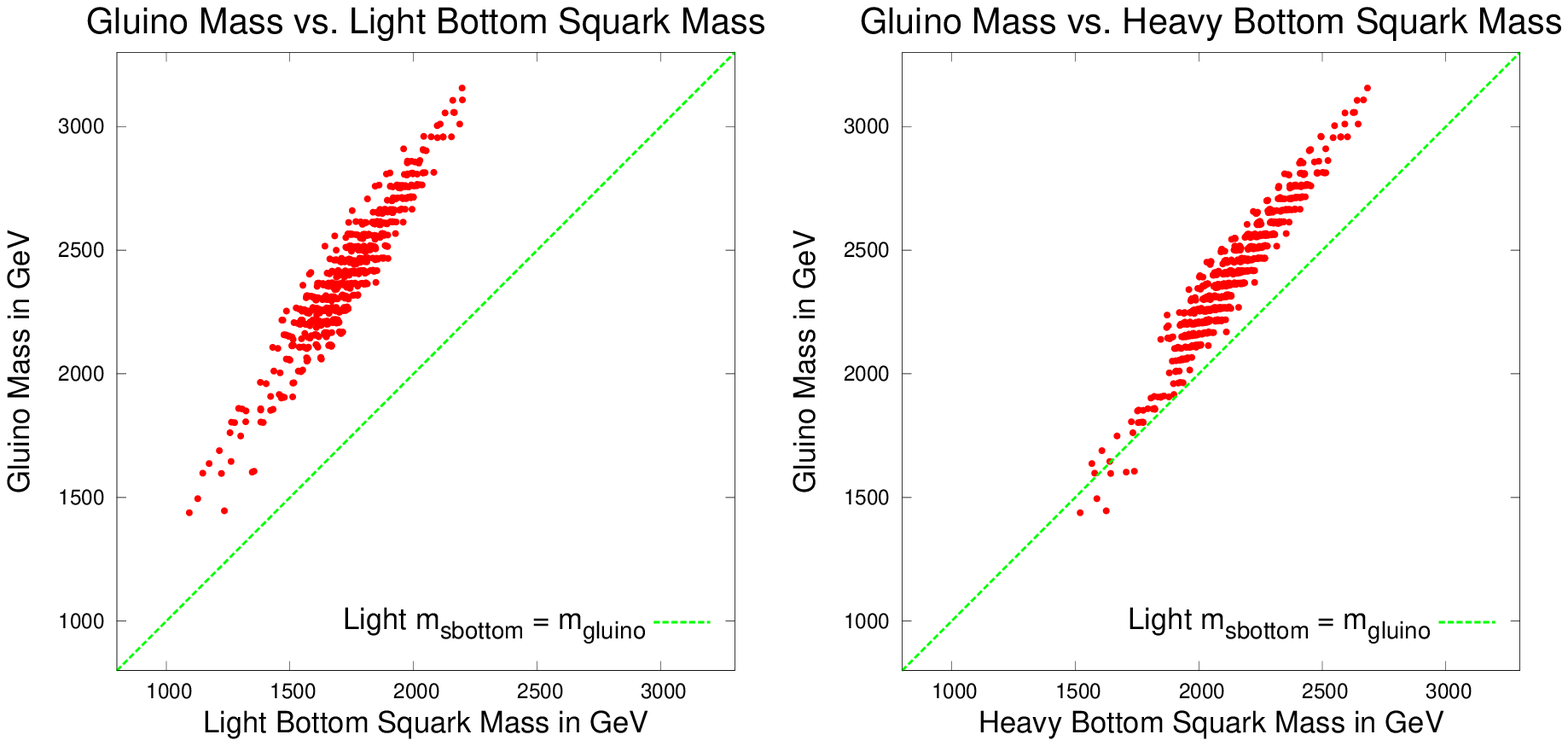}
\caption{These scatter plots provide a comparison between the bottom squark 
masses and the gluino mass in the models found by the scan.}
\label{delta m sbottom}
\end{figure}

\begin{figure}
\centering
\includegraphics[height=2.8in]{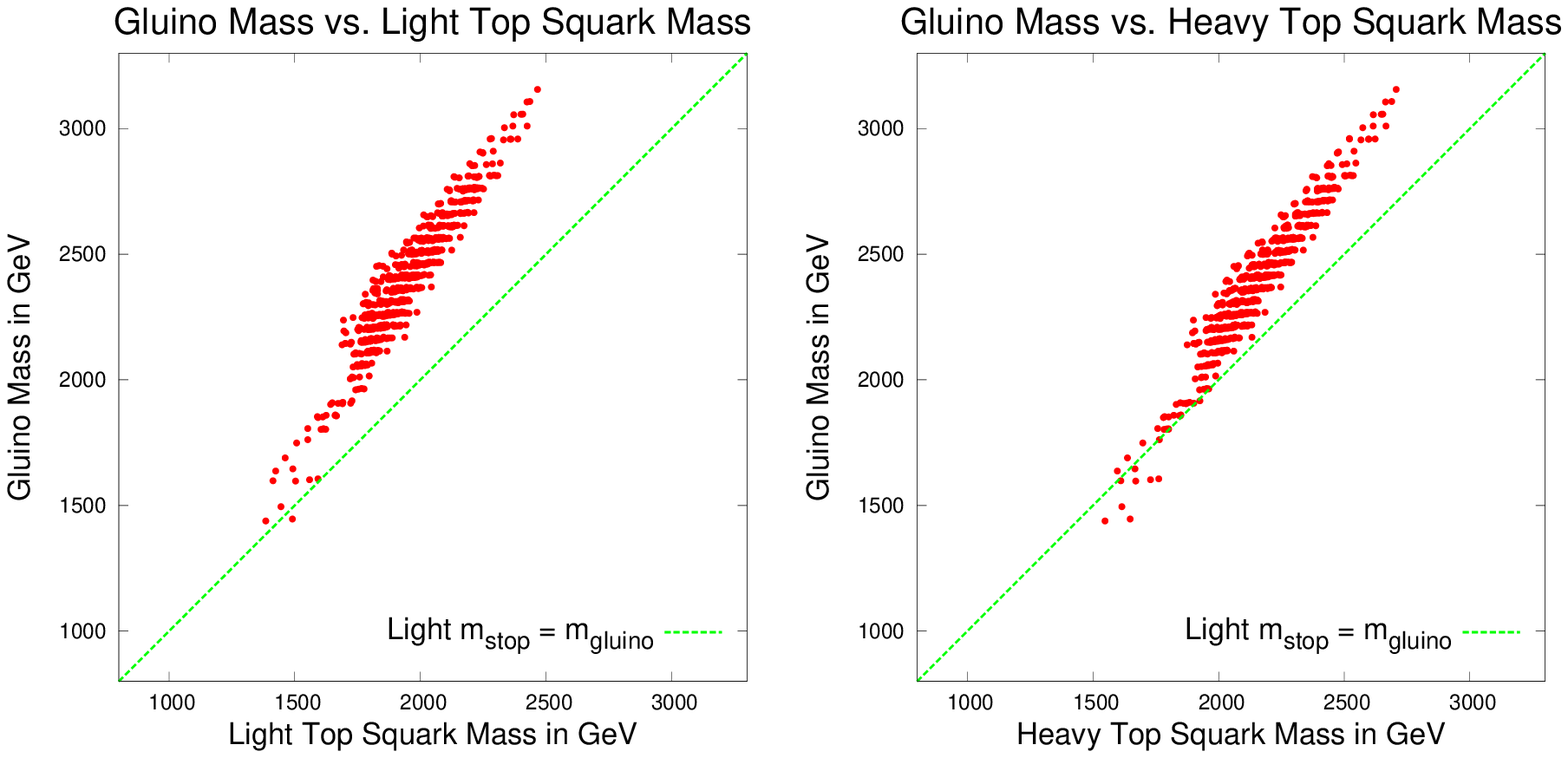}
\caption{These scatter plots provide a comparison between the top squark 
masses and the gluino mass in the models found by the scan.}
\label{delta m stop}
\end{figure}

In all models found by this scan, the light sbottom is the lightest squark and is lighter than the gluino. The relative lightness of this (mostly right-handed) sbottom (though it is not lighter than $\sim 1$ TeV) is due both to the effect of large Yukawa couplings on the sbottom mass RGE and the effect of D-terms, which lower the mass of right-handed down-type squarks when $M_D^2 > 0$ (as we demand in the scan). The lightest stop is also light in our models; generally lighter than the gluino. In an appreciable fraction of the models in this model set, the heavier stop and sbottom squarks are also lighter than the gluino. First and second generation squarks, however, tend to be somewhat heavier than the gluino, as is shown in Fig.~\ref{squarks}.

\begin{figure}
\centering
\includegraphics[height=2.8in]{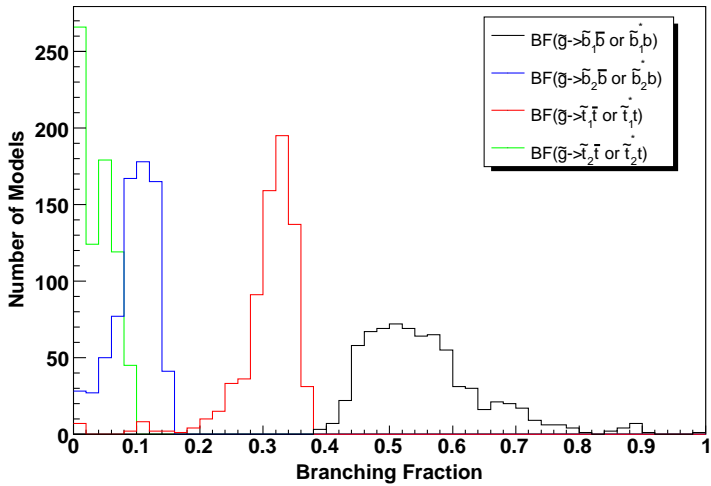}
\caption{This histogram shows the distribution of branching fractions of the gluinos to various third generation squark mass eigenstates and the associated quarks in the models found by the scan.}
\label{bf}
\end{figure}

Since the light sbottom (and generally the light stop and usually the other third-generation squarks) are lighter than the gluino, the dominant decay of the gluino will be two body decays to either sbottom and bottom quark or stop and top quark.  The branching fractions for decays to these squarks and quarks are shown in Fig.~\ref{bf}. The branching fractions were obtained using \texttt{SUSY-HIT 1.3}~\cite{Djouadi:2006bz}.

The stops and sbottoms produced in gluino decays will in turn decay to charginos and neutralinos, including the Higgsino-like charginos and neutralinos due to the large bottom and top Yukawas.  Then, the charginos and neutralinos will decay either to the LSP and an additional Higgs or electroweak gauge boson or to states with taus, as the stau is generally the Next to Lightest Supersymmetric Particle (NLSP). Thus the primary signatures of the SUSY scenario considered here are cascade decays, with an enriched fraction of b-jets, boosted tops, and taus. Clearly this model is discoverable at the $14$ TeV LHC.  
However, at the $7$ TeV LHC it will be very difficult to discover sparticles in the standard channels.  
In particular, the inclusive gluino pair production cross section in the model with the lightest gluino
found by the scan ($1438$ GeV) was investigated at NLO using \texttt{Prospino~2.1}~\cite{Beenakker:1996ch}. 
The gluino pair production cross section for this model is  $\sim 0.07$ fb, 
while the associated squark gluino production cross section is $\sim 0.09$ fb.
This suggests that the production cross section for these processes is generically small in this scenario at the $7$ TeV LHC.

\subsection{Dark Matter}

\begin{figure}
\centering
\includegraphics[height=2.8in]{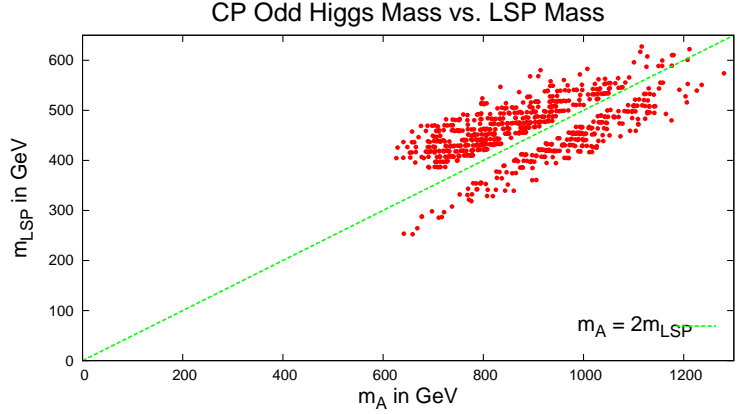}
\caption{This scatter plot shows the distribution of the CP-odd Higgs mass $m_A$ and the mass of the neutralino LSP. That the ratio between the two is generically close to $2$ suggests that neutralino annihilation through the CP-odd Higgs is efficient in all of the models selected by the scan.}
\label{cp odd ratio}
\end{figure}

\begin{figure}
\centering
\includegraphics[height=2.8in]{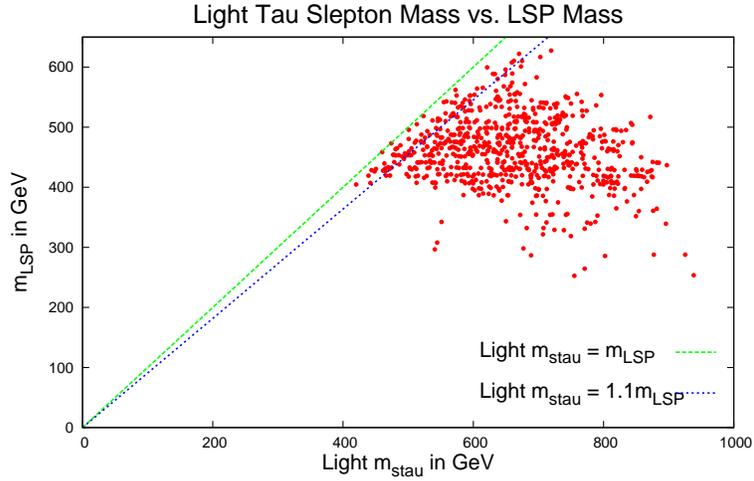}
\caption{This scatter plot shows the the distribution of the lighter $\tilde{\tau}$ slepton mass and the neutralino LSP mass. For models with low values of this quantity ($m_{\tilde{\tau}}/m_{LSP}\lesssim 1.1$), stau coannihilation may also play an important role in bringing the thermal LSP relic density down to the WMAP value.}
\label{stau ratio}
\end{figure}

In the scenario considered here, the neutralino LSP can have a thermal relic density in agreement with WMAP.  This is largely due to LSP annihilation through the CP-odd Higgs.  As can be seen in Fig.~\ref{cp odd ratio}, $m_A$ is roughly $2m_{\text{LSP}}$ in all of the models selected by the scan. As noted above, the lightest stau is often the NLSP; the ratio of the light stau mass and the LSP mass is shown for the models found by the scan in Fig.~\ref{stau ratio}. We note from the distribution of this ratio that there are models found by the scan for which stau coannihilation will be important (those where the mass ratio is $\lesssim 1.1$), however in most models stau coannihilation becomes negligible.

\begin{figure}
\centering
\includegraphics[height=2.8in]{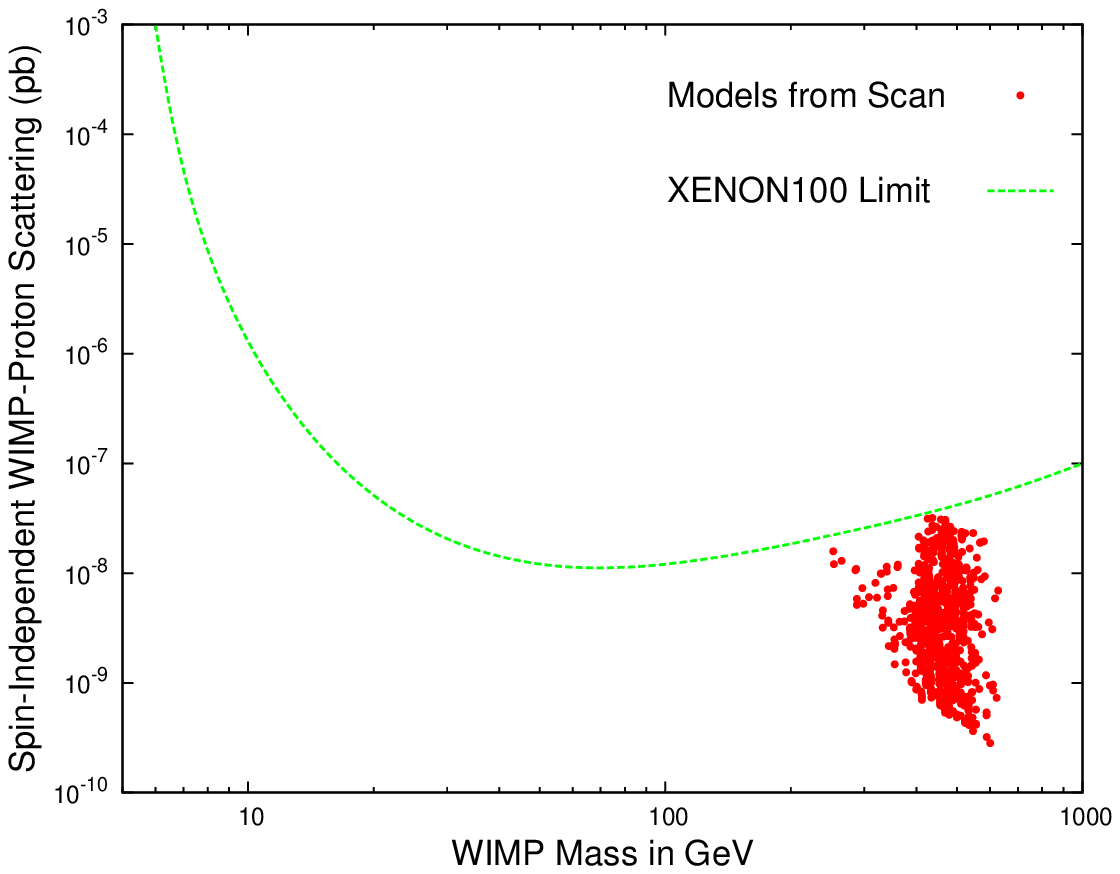}
\caption{Here we show the WIMP mass and WIMP-proton scattering cross section for models found in the scan as compared to the XENON100 limits on the spin-independent WIMP-nucleon cross section.}
\label{wimp}
\end{figure}

We have already noted that efficient annihilation through the CP-odd Higgs and (in some cases) coannihilation with the light stau allow the neutralino relic density to be consistent with WMAP values. This is different than the case in more standard Yukawa Unification scenarios, e.g.\cite{Baer:2008jn,Baer:2008yd}, where the overproduction of bino-like neutralinos in the early universe requires one to introduce a new state, such as an axino, as the true LSP. Fig.~\ref{wimp} shows the LSP mass and WIMP-nucleon cross section as compared with the XENON100 bounds. We note that the models found by the scan, which have the lowest values of spin-dependent WIMP-nucleon cross section, generally have higher 
LSP masses. Conversely models found by the scan with lighter LSPs should be discovered or ruled out by the next generation of direct detection experiments. Direct detection constraints were important in setting the lower bound on LSP masses found by the scan of $\sim 250$ GeV.

\section{Conclusions}\label{conclusions}

We have presented a scenario which adds to the framework of Yukawa-unified $SO(10)$ SUSY GUTs a mixing between generations in the charged lepton sector, which produces neutrino mixing. We find that the specific values of this mixing angle that one obtains from the values of the top, bottom, and tau Yukawa couplings at the GUT scale, in the specific $SO(10)$ inspired parameter space considered, suggest $\sin^2{2\theta_{23}} \approx 1$, in accordance with data, in the limit where no other large mixings are present.

Thus, while this scenario adds a parameter to the framework, the natural value taken by this parameter explains observed physics. Another property of this scenario is that we find viable points in our parameter space with sparticle masses above the current bounds, and quite generally above the reach of the $7$ TeV LHC, but likely within the reach of the $14$ TeV LHC. Viable points in our parameter space also have neutralino dark matter in agreement with WMAP and direct dark matter detection search constraints. Finally we find viable points in our parameter space for which the SM-like Higgs mass is between $118$ and $122$ GeV; a prediction which the $7$ TeV LHC will test. 
In future work, we will consider the collider and flavor phenomenology of this 
scenario in more detail and study the importance of the
assumption of small neutrino Yukawa couplings made in this work, as well 
as the effects of considering all three generations.
We look forward to tests of this scenario at the LHC.

\section{Acknowledgments}\label{Acknowledgments}

We would like to acknowledge useful conversations with Arjun Menon, Pedro Schwaller, and Lian-Tao Wang. This work is supported in part by the U.S. Department of Energy under contract numbers
DE-AC02-06CH11357, DE-FG02-91ER40684, and DE-FGO2-96-ER40956.

\end{document}